# SARAS CD/EoR Radiometer: Design and performance of the Digital Correlation Spectrometer

B. S. Girish[1], K. S. Srivani, Ravi Subrahmanyan, N. Udaya Shankar, Saurabh Singh, T. Jishnu Nambissan, Mayuri Sathyanarayana Rao, R. Somashekar and A. Raghunathan

*Raman Research Institute, C. V. Raman Avenue, Sadashivanagar, Bangalore 560080, India*

bsgiri@rri.res.in

In the currently accepted model for cosmic baryon evolution, Cosmic Dawn (CD) and the Epoch of Reionization (EoR) are significant times when first light from the first luminous objects emerged, transformed and subse- quently ionized the primordial gas. The 21-cm (1420 MHz) hyperfine transition of neutral hydrogen, redshifted from these cosmic times to a frequency range of 40 to 200 MHz, has been recognized as an important probe of the physics of CD/EoR. The global 21-cm signal is predicted to be a spectral distortion of a few 10's to a few 100's of mK, which is expected to be present in the cosmic radio background as a trace additive component. SARAS (Shaped Antenna measurement of the background RAdio Spectrum) is a spectral radiometer purpose designed to detect the weak 21-cm signal from CD/EoR. An important subsystem of the radiometer, the digital correlation spectrometer, is developed around a high-speed digital signal processing platform called pSPEC. pSPEC is built around two quad 10-bit analog-to-digital converters (EV10AQ190) and a Virtex 6 (XC6VLX240T) field programmable gate array, with provision for multiple Gigabit Ethernet and 4.5 Gbps fibre optic interfaces. Here we describe the system design of the digital spectrometer, the pSPEC board, and the adaptation of pSPEC to implement a high spectral resolution (61 kHz), high dynamic range ($10^5$:1) correlation spectrometer covering the entire CD/EoR band. As the SARAS radiometer is required to be deployed in remote locations where terrestrial radio frequency interference (RFI) is a minimum, the spectrometer is designed to be compact, portable and operating off internal batteries. The paper includes an evaluation of the spectrometer's susceptibility to RFI and capability to detect signals from CD/EoR.

## 1. Introduction

Cosmic Dawn (CD) and the Epoch of Reionization (EoR) represent significant moments in the history of the evolving universe when growing matter inhomogeneities transformed the primordial gas to one with galaxies surrounded by a diffuse intergalactic medium. There is considerable uncertainty in our understanding of the precise timing of events across this period and of the nature of the first stars and first ultra-faint galaxies that lit up the universe and transformed the state of the gas; the uncertainty is owing to lack of observational constraints. It has long been recognised that the 21-cm (1420 MHz) hyperfine transition of neutral hydrogen, redshifted from this interval in cosmic time to a frequency range of about 40 to 200 MHz, could potentially serve as an important probe of CD/EoR. Two complimentary approaches have been pursued to probe the evolving hydrogen in CD/EoR via redshifted 21-cm: one, using a radiometer (an antenna connected to a spectrometer) to measure the global spectrum of the radio sky and hence the mean evolution of the hyperfine level populations and ionisation state; second, using large interferometer arrays to measure the fluctuations in 21-cm brightness temperature at those epochs. Both approaches require careful design of radio telescopes so that confusing spectral structure from unwanted additives and bandpass calibration errors are minimised.

| Experiment | Operating Frequency range MHz | Sampling frequency MSps | ADC bit precision | Spectral Resolution kHz |
|---|---|---|---|---|
| SARAS | 40-250 | 500 | 10 | 61 |
| EDGES | 50-190 | 400 | 14 | 6.1 |
| LEDA | 30-88 | 196.6 | 8 | 24 |
| PRIZM | 0-250 | 500 | 8 | 61 |

Table 1. Spectrometer specification for SARAS compared with other ongoing experiments attempting to detect global redshifted 21-cm from CD/EoR.

SARAS is a radiometer experiment attempting to measure the global redshifted 21-cm from CD/EoR. The work presented here is on the design and performance of a digital correlation spectrometer purpose built for SARAS. Apart from SARAS (Singh S. *et al.*, 2017, 2018a,b), there are a few other global 21-cm experiments; for example, EDGES (Monsalve R. A. *et al.*, 2017a,b; Bowman *et al.*, 2018), LEDA (Bernardi *et al.*, 2016; Price *et al.*, 2018), and PRI²M (Philip *et al.*, 2019). Interferometer arrays that are operational and have been built with a key science goal of detecting the 21-cm power spectrum are MWA (Lonsdale *et al.*, 2009), LoFAR (van Haarlem *et al.*, 2013) and HERA (DeBoer *et al.*, 2017). We present in Table 1 a comparison of digital receiver specifications for global EoR experiments underway, including the SARAS digital receiver that is described in this paper.

Recent advances in processing capabilities of Field Programmable Gate Arrays (FPGAs), combined with high-speed analog-to-digital converters (ADC), make the implementation of high-resolution, high dynamic range spectrometers for detecting 21-cm from CD/EoR possible. ADCs capable of directly sampling signals in the frequency band of interest for CD/EoR–the 40 to 200 MHz radio-frequency (RF) band–without the need for nonlinear analog down conversion, and modern FPGAs with their high-density programmable features, have enabled development of robust, high-performance RFI-tolerant CD/EoR spectrometers with better control over systematics.

The SARAS system configuration has evolved over the years, from the first version that had a fat-dipole antenna, to SARAS 2 and now SARAS 3 that have monopole antennas. In all versions, the sky signal is split at some stage, band limited in parallel analog receiver chains to below 250 MHz, before being presented to a digital correlation spectrometer. The correlation spectrometer concept has been a feature of all versions of SARAS, along with phase switching between the two arms so as to cancel any additive systematics that couple into the receiver chains. At the core of the SARAS digital spectrometer is a high-speed signal processing platform, pSPEC (precision SPECtrometer), consisting of two 10-bit ADCs from e2V technologies followed by a Virtex-6 FPGA. The ADCs, operating at 500 MSps, digitise the two analog signals. In the FPGA, blocks of ADC samples from each receiver chain are weighted with a window function and separately Fourier transformed. The averaged power spectra corresponding to each of the two arms, as well as the complex cross power spectrum corresponding to the product, are computed and streamed out to an acquisition computer through a Gigabit Ethernet interface.

The SARAS radiometer consists of an antenna, antenna-base electronics that is located in an enclosure immediately beneath the antenna, an analog signal conditioning unit and a digital receiver. The analog signal conditioning unit and digital receiver are located about 100 m away from the antenna. The antenna base electronics is designed to cycle through a number of switching states to present successively the antenna and cold & hot terminations to the receiver for calibration. The digital receiver enclosure includes the digital spectrometer and a laptop computer and electronics for generating control signals for the antenna base electronics; the digital receiver controls the radiometer operations, generating signals for switching the antenna base electronics, synchronously controlling the digital signal processing in the spectrometer and also the acquisition into the laptop computer. Signal transport of digital control signals to the antenna and of analog signals from the antenna base to the analog signal conditioning unit is via optical fiber.

In the following sections, we present the system description, including the rational behind choice of components, integration of various sub-sections of the digital correlation spectrometer, packaging, thermal management and, finally, laboratory tests carried out to evaluate the SARAS spectrometer.

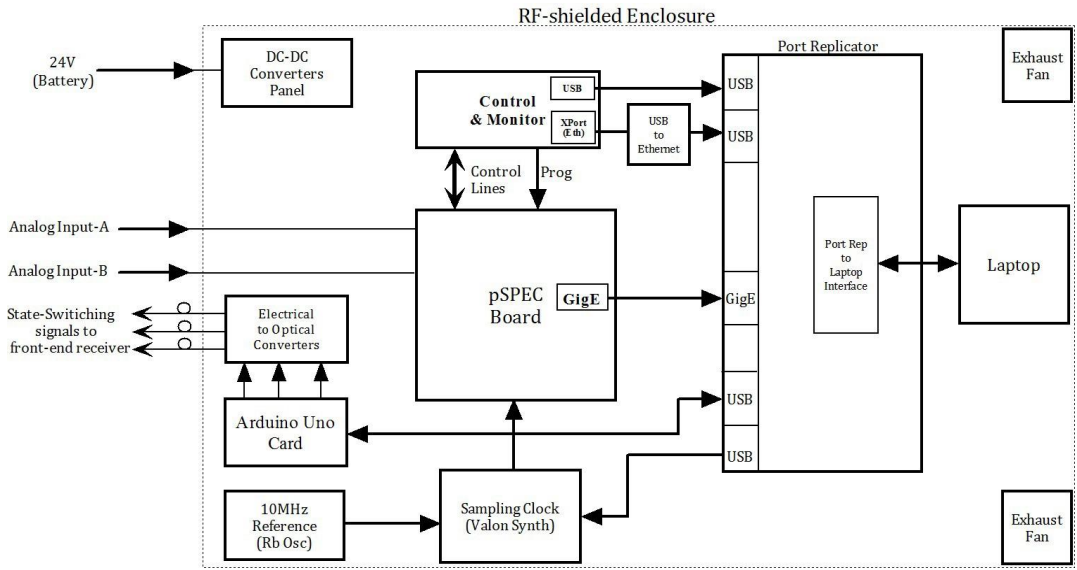

Fig. 1. System level architecture of the SARAS Spectrometer.

## 2. System Description

Design of any digital correlation spectrometer for detecting EoR, which is an extremely weak signal embedded in orders of magnitude brighter foreground, is challenging. The theory of the formation of First Stars and the subsequent heating and reionization of the gas by First Light that emerges from the earliest ultra-faint galaxies is highly uncertain; therefore, the predicted 21-cm signal is largely unknown but nevertheless expected to be in the range of a few 10's to a few 100's of mK brightness temperature. Since the sky foreground brightness in the frequency band of interest is in the range of several 100's to 1000's of K, the dynamic range of spectral receivers aiming to detect the global 21-cm signal needs to be about a million to one. Additionally, since the receiver band includes FM and TV channels, which can be several orders stronger than the foregrounds, it is required to operate in the presence of substantial terrestrial RFI. Every sub-system in the signal path, from the sensor of the EM field to the back-end digital spectrometer, needs careful design/engineering to adequately suppress the deleterious effects of RFI.

Fig. 1 shows the system architecture of the SARAS spectrometer. At the core of the spectrometer is the pSPEC board, built around two, four-core time-interleaved Analog-to-Digital Converters (ADCs) and a Virtex-6 FPGA. For wideband applications in radio astronomy, FPGAs are preferred due to the large number of high bandwidth input/output (I/O) pins that may be used to stream wideband data sampled at high rates to implement computationally-intensive signal processing algorithms. FPGAs, with them in-built specialised hardware blocks, are well suited to demultiplex high-speed serial data into multiple parallel data streams, enabling a parallelised channelisation structure to transform the samples of a wideband time-domain signal into multiple narrow sub bands for further processing.

The digital spectrometer has a laptop computer for data acquisition and control. The laptop connects to a docking station that provides power to the laptop and has a port replicator with USB and Gigabit Ethernet ports. Through these ports, the laptop connects to the pSPEC board and also to peripherals like a control and monitor card, a synthesizer and an Arduino card, which is used to generate the state-switching signals to the RF front-end receiver of SARAS. A 10 MHz signal derived from a rubidium oscillator provides a stable reference to the synthesizer used to generate the sampling clock to the ADCs. The entire digital receiver is assembled inside an RF-shielded enclosure to suppress broad band and narrow band interfering signals, which are inevitably generated in the digital hardware, from propagating to the antenna and thereby limiting the sensitivity of the radiometer via self-generated RFI.

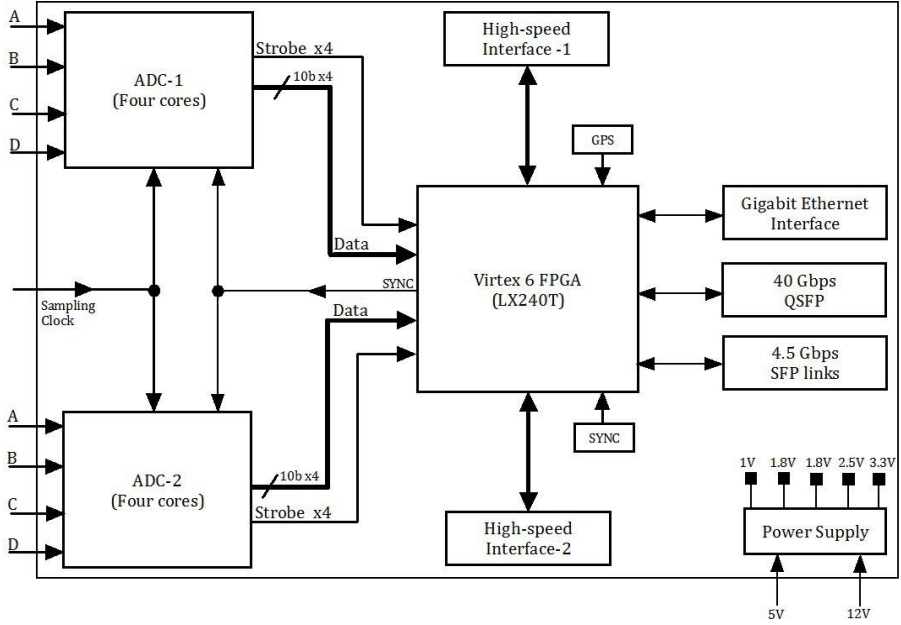

Fig. 2. Block diagram of the pSPEC architecture.

It may be noted here that pSPEC was envisaged as a common DSP platform catering to the requirements of CD/EoR (40–200 MHz) and also for detection of cosmological recombination lines (APSERa;(Sathyanarayana Rao *et al.*, 2015)), where a wide-band 2 GHz bandwidth correlation spectrometer is required. In this paper, we report the adaptation of pSPEC in the development of a high-precision, high dynamic range digital correlation spectrometer to detect the global CD/EoR signal.

## 2.1. The pSPEC board

The precision spectrometer (pSPEC) board is a Xilinx FPGA based signal processing platform and is the primary building block of the digital correlation spectrometer. Fig. 2 shows the major components within the pSPEC board. pSPEC features two 10-bit quad time-interleaved ADCs (ADC-1 and ADC-2; both are EV10AQ190CTPY from e2V technologies), a Virtex-6 LX240T FPGA from Xilinx, two high-speed electrical interfaces, a Gigabit Ethernet interface and optical interfaces consisting of eight 4.5 Gbps Small Form-factor Pluggable (SFP) ports and a 40 Gbps Quad Small Form-factor Pluggable (QSFP) port. Additionally, there are dedicated connectors to accept "SYNC" signals to control the synchronous sampling at the two quad ADCs and a Global Positioning System (GPS) signal to time-stamp data streaming out of the pSPEC. Fig. 3 shows a photograph of the pSPEC board that has been implemented as a mixed-signal printed circuit board with a height of 233 mm (6U standard) and a depth of 257 mm.

### 2.1.1. The Analog-to-Digital Converter in pSPEC

The choice of the ADC for the pSPEC board was driven by the requirement of developing a common signal processing platform capable of digitizing analog signals having bandwidths ranging from a few hundred MHz to about 2 GHz. A market survey of candidate broadband data converters possessing 2 or more channels with at least 10 bits of resolution resulted in the choice of EV10AQ190. It is a 10-bit quad core device, with each core having a maximum sampling rate of 1.25 GSps. A summary of the features and expected performance of the ADC used in pSPEC is in Table 2.

The device has an on-chip programmable register that allows configuring the ADC in a mode in which it samples four independent analog channels, or samples two channels using a pair of ADC cores per channel in time-interleave mode, or samples a single channel by time interleaving between all four cores; the three modes provide maximum sampling rates of 1.25 GSps, 2.5 GSps and 5 GSps respectively. Thus, the two

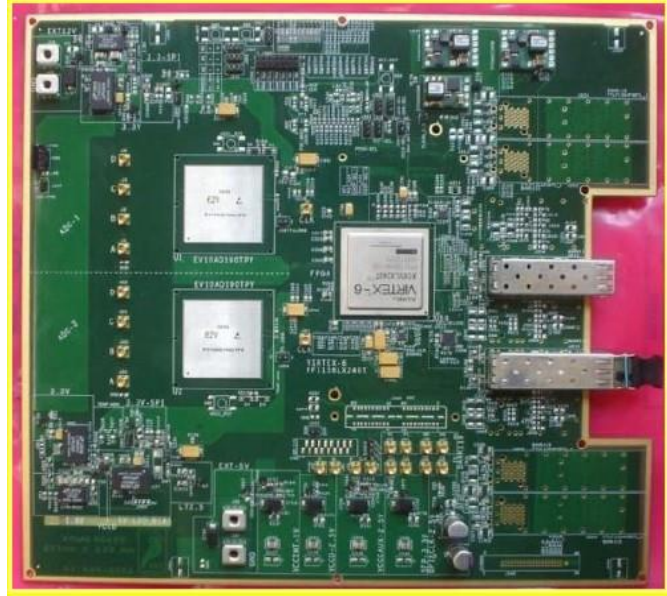

Fig. 3. Photograph of the pSPEC board.

| |
|---|
| Quad core ADC, each with 10-bit resolution |
| On-chip programmable registers of ADC are accessed via SPI bus Analog input bandwidth of 3.2 GHz |
| Full-scale input range of 500 mVp-p ($\sim -2$ dBm) |
| SNR of 52 dB for Fin=100 MHz and Fclk=1.25 GSps (independent channel mode) |
| ENOB of 8.3 for Fin=100 MHz and Fclk=1.25 GSps (independent channel mode) Power dissipation of 1.4 W per channel |
| 380-pin EBGA package |
| Isolation between channels: 60 dB |

Table 2. Specifications of the EV10AQ190 ADC used in pSPEC.

ADCs on pSPEC enable digitization of 8 independent analog signals (each up to 625 MHz bandwidth) or a pair of wideband analog signals (each up to 2.5 GHz bandwidth) in time-interleaved mode. The 10-bit wide data from the individual ADC cores are output on a low-voltage differential signaling (LVDS) bus using double data rate (DDR) interface, permitting the latching of data on rising and falling edges of the ADC data strobe signal, at maximum rate of 625 MHz. For the SARAS spectrometer, the ADCs are configured in independent channel mode of operation and two ADC cores, each of them clocking at 500 MHz rate, are used to digitize a pair of baseband analog signals each of which have a bandwidth of about 250 MHz. When a multi-channel ADC is used to digitize multiple analog signals for a correlation spectrometer, crosstalk between channels is a major concern, especially in precision and high dynamic range applications. In the pSPEC board, we have provided adequate spacing between routes and utilized multiple routing layers within the 18-layer PCB of pSPEC; as a consequence, within each ADC, 58 dB isolation between cores has been achieved. This compares favorably with expectations based on the device data sheet, which specify about 60 dB isolation.

The signal-to-noise ratio (SNR) of an ADC is related to the effective number of bits (ENOB) $b$ by (Kester, 2009)

$$SNR = (6.02b + 1.76)\ dB. \tag{1}$$

From Equation 1, the signal-to-noise ratio (SNR) of an ideal 10-bit ADC is 62 dB, assuming that the ENOB is 10 bits. The specifications of the ADC suggest an ENOB of 8.3, which implies that the ADC device used in pSPEC may be expected to provide sampling with a SNR of 51.7 dB.

One of the important factors that determines the performance of a high-speed, high-resolution ADC is the quality of its sampling clock. The degradation in the SNR of the ADC due to any jitter in the sampling clock is given by

$$SNR = -20\log(2\pi f_o \tau) \qquad (2)$$

where $f_o$ is the highest frequency component in the analog signal that is input to the ADC and $\tau$ is the jitter in the sampling clock. From Equation 2, for $f_o$ of 250 MHz and assuming an ideal SNR of 62 dB, allowable jitter in the sampling clock should be less than 0.5 ps so as not to degrade performance beyond ideal device limitations.

The sampling clock for pSPEC is generated using a programmable frequency synthesizer Valon 5008 from Valon Technologies. It has the provision to take in an external stable reference signal so as to provide better stability and spectral purity for the generated sampling clock. In SARAS, the primary frequency standard for deriving the sampling clock is a rubidium oscillator. The 10 MHz signal from the rubidium oscillator is used as reference for the Valon synthesizer to derive the 500 MHz sampling clock for the ADCs on the pSPEC board; use of such an ultra-stable reference provides a clock with jitter limited to 2 fs, which is substantially below the 0.5 ps requirement.

Preceding the digitizer, the voltage gains in the analog receiver chain of SARAS are adjusted such that the total power at the analog input of the ADC is about −28 dBm – arrived at since the ADC'S total harmonic distortion is minimum– when an ambient temperature load is connected to the receiver. Assuming that no other noise is present apart from the quantization noise arising from the quantization step of 0.488 mV, the quantization noise power referred to the ADC input is about −65 dBm, for an ideal 10-bit ADC; assuming that the ENOB of the pSPEC ADC is 8.3, the quantization noise power referred to the ADC input is −54.7 dBm. Thus the quantization noise increases the system temperature by less than 0.5%.

The pSPEC ADC clips a sinusoidal signal if its power exceeds 2 dBm (full-scale input). The total power of −28 dBm in any Gaussian random signal presented at the analog input of the ADC is 26.7 dB greater than the quantization noise of the ADC, and also 26 dB smaller than the full-scale range of the ADC. This ensures that a headroom corresponding to about 4 bits is available to accommodate any strong RFI that may be substantially greater that the total band power from the sky and receiver noise.

### 2.1.2. The FPGA used in pSPEC

The FPGA on pSPEC was envisaged to grab data from all eight ADCs, deserialize them into parallel data streams that can be clocked within the realm of Virtex 6 FPGA technology, and transform data that is in time sequence to spectral domain. The channel data is then processed in the FPGA to correlate between data streams of the different ADCs, and transport the averaged correlation computed in channelized data to a computer for further processing and recording on to hard disk.

There were a number of considerations leading to the choice of an FPGA for pSPEC. A requirement was that the FPGA have adequate I/O pin-count along with inbuilt deserializer to handle the data bits that would stream at 625 Mbps (double data rate) from all eight ADCs of pSPEC. Additionally, the FPGA was required to have logic and specialized resources like a block RAM (BRAM), DSPE1 slices that would be required for implementation of a real-time FX correlation spectrometer, and embedded blocks like Tri-mode Ethernet Media Access Controllers (TEMAC) for control and data transfer. The cost of the FPGA device was also a consideration. These led to XC6VLX240T-FF1156 as the FPGA for the SARAS spectrometer. The selected Virtex-6 FPGA device is a 1156-pin flip-chip fine-pitch ball grid array (BGA) package built using 40 nm technology and belongs to the sub-family LXT that is meant for high-performance logic with advanced serial connectivity.

XC6SX315T, which is an FPGA that has a footprint compatibility with XC6VLX240T, has also been chosen for pSPEC as an alternate configuration. This FPGA is specifically meant for pSPEC cards for APSERa in which all the eight ADC cores may be used in time-interleaved mode and clocked at 1 GSps. This Virtex-6 SXT sub-family FPGA comes with enhanced DSP48E1 slices, BRAM and logic resources.

Both FPGAs have 600 user I/O pins sufficient to interface all eight ADCs and still have provision for general-purpose I/O lines and high-speed electrical interface between pSPEC boards. Specifications of the two FPGAs selected for pSPEC are summarised in Table 3.

| Parameter | LX240T FPGA | SX315T FPGA |
|---|---|---|
| Speed grade | -2 | -2 |
| Logic cells | 241152 | 314880 |
| Slices | 37680 | 49220 |
| Distributed RAM (kb) | 3650 | 5090 |
| Number of 36 kb BRAM | 416 | 704 |
| Multi-mode clock manager | 12 | 12 |
| Maximum single-ended I/O | 600 | 600 |
| Maximum differential I/O pairs | 300 | 300 |
| DSP48E1 slices | 768 | 1344 |
| Ethernet MACs (TEMACs) | 4 | 4 |
| GTX transceivers (up to 6.6Gbps) | 24 | 24 |
| PCI Express interface blocks (SR 2.0) | 2 | 2 |

Table 3. Specifications of the Virtex 6 FPGAs on pSPEC

The 1156-pin BGA package of both LX240T and SX315T FPGAs are organized in a matrix of 34 rows by 34 columns. The 380-pin package of the ADC is organized in a matrix of 24 rows by 24 columns. At least 6 signal routing layers (Note , 2010) are required to achieve full breakout of all the 600 I/O pins available on the FPGA package. Based on the component density, the pin-counts of the selected FPGA and ADC, and the need to provide for high-speed optical and electrical interfaces, pSPEC is laid out on an 18-layer printed circuit board (PCB) consisting of eight signal routing layers and ten power and ground planes. To maintain signal integrity of high frequency analog signals and high speed digital routes, the two outer layers of the 18-layer pSPEC PCB are fabricated using low loss tangent, low coefficient of thermal expansion (CTE) and high glass transition temperature laminate EM-827 from Elite Material Corporation.

## 2.2. *Digital signal processing in pSPEC*

Inside the Virtex 6 FPGA of pSPEC, the correlation spectrometer is realized as four distinct stages: grabbing of ADC data samples, weighting of data samples, Fourier transformation using a 16384-point FFT engine (F-engine) and lastly a multiply-and-accumulate stage (X-engine). The F-engine is implemented as a split-FFT (MxN point) architecture.

Fig. 4 shows the block diagram of the firmware architecture for the SARAS spectrometer. Analog signals from the pair of analog signal processing chains are fed to two ADC cores for digitization. The ADCs on pSPEC are supplied with a low-jitter sampling clock of 500 MHz from a Valon 5008 synthesizer. As mentioned above, a 10 MHz signal derived from a rubidium oscillator forms the primary frequency standard and is provided as an external reference to this synthesizer. At the FPGA interface, individual bits of each ADC are grabbed and deserialized by a factor of four using input serializer-deserializer (ISERDES) primitives. The ADC data stream is required to be demultiplexed only by a factor of two to bring the clock frequency of the two parallel paths within the realm of the Virtex 6 FPGA. However, for DDR interface, the minimum deserialization factor of four in the ISERDES primitives constrains the design to demultiplex the data stream by the same factor. Subsequently, to keep the FPGA resource consumption to a minimum while implementing the correlation spectrometer firmware, we have used a data buffer to reduce the number of demultiplexed paths from four to two.

Prior to Fourier transformation, the demultiplexed samples are weighted with appropriate coefficients of the minimum 4-term window function (Nuttal, 1981). Considering the high sensitivity requirement for observing the EoR signal, windowing helps to limit the spill over from any strong narrow-band interfering signal to adjacent spectral channels. The minimum 4-term window function used in pSPEC provides, ideally, a sidelobe suppression of about 98 dB; in practice, due to the 10-bit digitization of the analog signals and the finite-precision representation of signals inside the FPGA, a sidelobe suppression of 80 dB is measured. The down weighting by the window function reduces the effective integration time by a factor of two and increases the noise-equivalent bandwidth of the channels by a factor of two; however, the severe windowing has the advantage that even with an RFI as strong as $10^6$ K, due to high suppression achieved, contamination of neighbouring and other spectral channels is limited to below 10 mK.

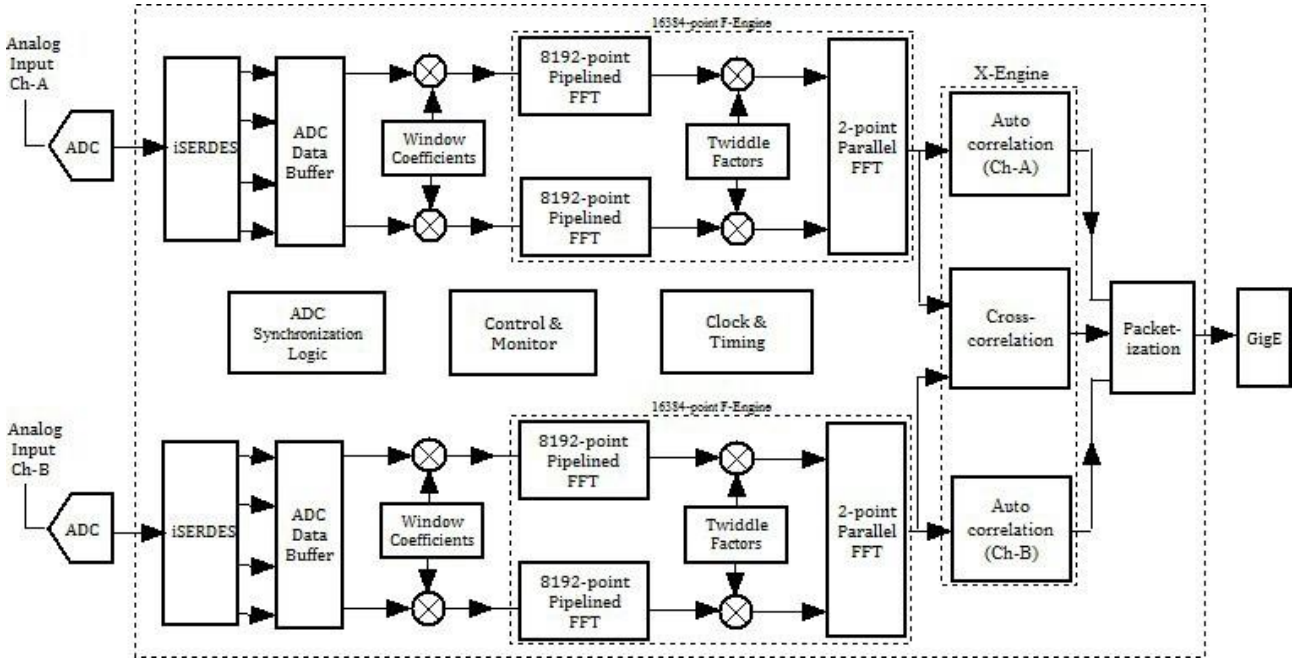

Fig. 4. FPGA firmware architecture of SARAS spectrometer.

The F-engine of the FX spectrometer consists of two 8192-point (N=8192) pipelined FFT cores operating in parallel, followed by two complex multipliers for phase rotation and a 2-point (M=2) parallel FFT to combine the outputs from the two parallel paths. While the 8192-point pipelined FFT cores are sourced from Xilinx IP-core generator, the 2-point parallel FFT is custom designed to compute a 2-point FFT in a single clock cycle. Using this scheme, a 16384-point streaming channelizer has been implemented, providing 8192 complex channels covering the band 0–250 MHz at the output. The 8192-point spectrum from the F-engine has spectral values spaced 30.518 kHz apart; however, the spectral resolution is 61.035 kHz owing to the windowing. The time required to compute each 16384-point FFT is 32.768 $\mu$s. Two such F-engines are instantiated inside the Virtex 6 FPGA to channelize the sampled signals from the two ADCs.

In the multiplier-integrator section (X-engine), blocks of 2048 FFT spectra are processed. For each of the two analog signal paths, separately, power spectra are computed from each of the complex FFTs and these are averaged to generate the auto-correlation power spectra corresponding to the signals in each of the paths. Simultaneously, a complex cross-correlation spectrum is computed by channel-wise complex multiplication of the FFT values from the two signal paths and averaging over the block of 2048 spectra. The choice of averaging time—over 2048 FFTs—avoids overflow in the multiply-and-accumulate resources of the FPGA and the processing of the block of 2048 spectra is realized in an on-chip integration time of 67 ms.

The three integrated spectra along with data markers and identifiers are packetized and streamed o u t of the FPGA at about 16 Megabytes per second using User Datagram Protocol (UDP). The on-chip integration time of 67 ms represents a tradeoff between the need for detection and mitigation of short time scale and transient RFI, and sustaining the data acquisition in real time without loss of data packets. UDP is used owing to its reduced bandwidth overhead and low-latency connection between applications. A dedicated hardware Tri-mode Ethernet Media Access Controller (TEMAC) block available inside the Virtex 6 FPGA, along with an external Small Form-factor Pluggable (SFP) module that provides a copper interface to the external world, is used for transferring data from pSPEC to the acquisition computer. 16 sets of integrated spectra are buffered within the pSPEC board, corresponding to a total integration time of about 1.072 s (67 ms $\times$ 16), and these are streamed out of pSPEC in a single data acquisition cycle. This allows the receiver to be switched between different system states, between Dicke switch positions, between calibration noise source in on and off conditions, etc. every 1.072 s and the 16 frames of data read out for each such state. The firmware for the operation of the FPGA was developed in VHDL using Xilinx ISE software for design flow and Mentor Graphics' ModelSim for HDL simulation. The optimised firmware operates at a clock frequency of 250 MHz, with less than 50% FPGA resource utilization. A summary of the FPGA resources utilized in implementing the FX correlation spectrometer is provided in Table 4.

| FPGA Resource utilization | Usage | Percentage |
|---|---|---|
| Number of occupied slices | 9478 out of 37680 | 25 |
| Number of RAM36E1 | 236 out of 416 | 56 |
| Number of MMCM | 6 out of 12 | 50 |
| I/O | 191 out of 600 | 31 |
| Number of DSP48E1s | 114 out of 768 | 14 |
| Number of TEMACs | 1 out of 4 | 25 |

Table 4. Summary of Virtex 6 LX240T resource utilization.

### 2.3. *Control and Data Acquisition*

While the pSPEC unit forms the heart of the digital spectrometer, a laptop computer that is part of the digital receiver is designated to be the master controller. It is used to configure and control the ADCs and FPGA on pSPEC through a Lantronix XPort and USB-based controller card (XPort card), control the real-time digital signal processing activities inside the FPGA, perform acquisition of integrated spectra streaming out of pSPEC via the Gigabit Ethernet interface, and synchronously control the switching of states of the analog electronics in the antenna-base receiver and the signal conditioning unit. The Valon synthesizer module and the Arduino Uno microcontroller card are interfaced with the laptop. The laptop computer selected to be the master controller is a Toshiba Portege R930 laptop, which is a portable laptop weighing 1.4 kg and with a 512 GB solid-state drive (SSD) for storage. In addition to being compact, durable and light-weight and relatively invulnerable to vibrations, SSDs, as compared to hard-drives, provide advantages like better system responsiveness and operating power efficiency as there are no moving parts.

The laptop has a docking connector to interface with a compatible Toshiba high-speed port replica- tor/docking station. The port replicator has a 10/100/1000 Mbps LAN interface, six USB interfaces and laptop eject lever with lock. When the laptop is docked, power to charge the laptop is provided through the DC-In port of the port replicator. With these features, the port replicator eliminates the inconvenience of connecting and disconnecting multiple cables whenever the laptop is required to be connected/disconnected from the spectrometer.

The digital spectrometer is wired in such a way that only when the laptop is docked in the port replicator and powered on, the Arduino Uno card receives its power and control-signals through a USB interface to the port replicator. A solid-state relay controlled by the Arduino Uno switches-in the +24 V to power all other units of the spectrometer.

The switch states of the analog receiver chain housed beneath the antenna and in the signal conditioning unit are controlled using three control signals generated from the combination of a program running on the laptop and the microcontroller on the Arduino Uno card. The control signals (i) switch the calibration noise source on/off, (ii) Dicke switch between antenna and an internal reference termination, and (iii) switch the output of the front-end receiver alternately between the two paths to implement phase switching in the correlation spectrometer. Switching the calibration noise source provides calibration data for correcting for the bandpass response of the receiver chain. Dicke switching rejects unwanted additives in the front-end receiver electronics. Phase switching the analog signal path is designed to reject common-mode unwanted additives that enter the signal path in the analog signal conditioning unit and samplers.

Four Broadcom optical transmitters, based on HFBR 1412Z low-cost 820 nm miniature link fibre optic components, mounted inside the digital spectrometer unit are used to convert the control signals (electrical) from the Arduino Uno card to optical. Four optical fibre cables connected to ST (housed) optical interfaces on the RF-shielded enclosure carry the control signals to the front-end receiver unit kept at a distance of about 100 m from the digital spectrometer unit. One of the four control signals is dedicated to remotely switch on the front-end receiver at the start of observations and automatically shut down both the analog front-end receiver and the digital receiver at the end of observing sessions, to save battery power. Optical fibres are preferred to metallic coaxial cables as they minimally affect the antenna behaviour. Synchronizing the switching of the states in the analog receiver via the control signals, and the operations of the pSPEC card and acquisition of averaged spectra corresponding to each of the states is carried out by a C program running on the laptop computer.

A setup script, written in C, routes the configuration bitstream of the FPGA through a USB interface between the port replicator and the XPort card. A high-speed cable connects the XPort card to the FPGA though one of the two high-speed interfaces. The programmable registers inside the ADCs are accessed through SPI interface between the FPGA and the ADCs. These registers are used to configure the ADC in one of the three operating modes, or write/read the Offset, Gain and Phase (OGP) control registers of each of the four cores in the ADC. The contents of the ADC registers are written to and read from a dedicated set of registers inside the FPGA. These registers in turn communicate with the laptop through the 10/100 Lantronix Ethernet interface on the XPort card.

A socket program is used to grab the UDP data packets streamed out of pSPEC. Spectrometer data corresponding to each of the switched states of the front-end receiver are extracted by stripping out data markers and identifiers from the acquired data and routed to designated buffers for further processing. Table 5 provides a summary of the specifications of the FX correlation spectrometer for SARAS.

| Spectrometer parameter | Value |
|---|---|
| Number of analog inputs | 2 |
| Sampled bandwidth | 250 MHz |
| Sampling speed | 500 MSps |
| ADC | EV10AQ190CTPY |
| ADC bit resolution | 10 (ENOB=7.8) |
| Nominal power at input to the ADC | $-28$ dBm |
| ADC headroom for RFI | 26 dB |
| Virtex 6 FPGA | XC6VLX240T (1156 pins) |
| Length of FFT | 16384 point |
| Precision in the FFT twiddle factors | 16,16 |
| Length of window function | 16384-point |
| Bit precision of coefficients of the window function | 18 bits |
| Maximum sidelobe levels in the spectral PSF | 81 dB |
| Spectral resolution following windowed FFT | 61.75 kHz |
| Time for computation of a single FFT | 32.768 $\mu$s |
| Number of on-chip spectral averages and integration time | 2048, 67.108ms |
| Data rate | 16 MBps |

Table 5. Specifications of the SARAS correlation spectrometer.

The digital spectrometer built around the pSPEC platform and associated sub-systems, including the laptop and port replicator, were assembled in a custom designed three-tiered metallic frame. The trolley is placed in a shielded enclosure. A picture of spectrometer with the door of the shielded enclosure open to show the frame with sub-systems is shown in Fig. 5.

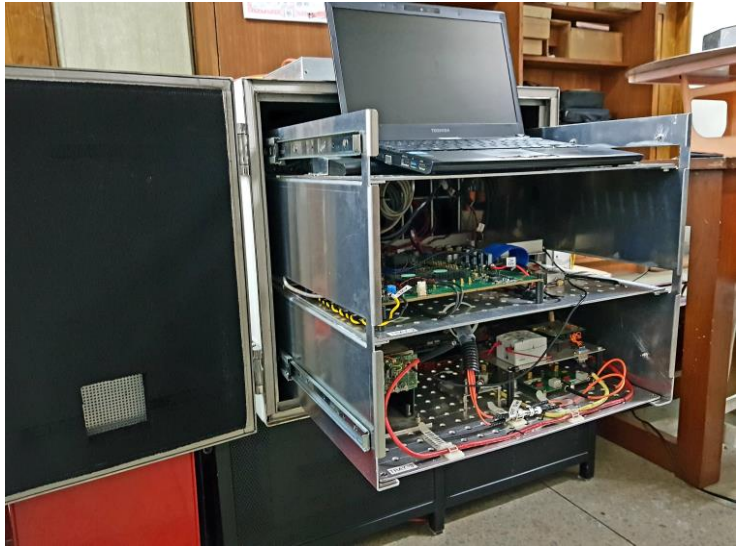

Fig. 5. Trolley showing various modules of SARAS 3 Digital Receiver

## *2.4. Shielded Enclosure for the digital spectrometer*

The global cosmological redshifted 21-cm signal is expected to be rather faint, with peak amplitude about 20–200 mK; therefore, terrestrial RFI may potentially confuse a detection. External RFI from communication and broadcast transmitters may be avoided by deploying the radiometer in remote sites where RFI power is adequately small. However, self-generated RFI from the receiver system needs to be prevented from coupling into the signal path so that it does not produce artefacts in the data and hence limit the sensitivity of the radiometer. As with any radio telescope, in SARAS the primary effort has been on pre- venting the clock frequency tones, their harmonics and wideband emissions from the high-speed digital electronics and data acquisition computer from reaching the antenna and feeding back into the system. Radio frequency shielding of these contaminating sources of emission by housing them in an RF-shielded enclosure helps in isolating them from the antenna and analog receiver chain.

The entire digital receiver of SARAS is housed inside a commercial RF-shielded enclosure which attenuates electromagnetic signals generated within the enclosure by about 75 dB in the CD/EoR band. To measure the isolation achievable from the enclosure, a radiating antenna connected to a sinewave generator was placed inside the enclosure to simulate a source of RF emission. A receiving antenna connected to a spectrum analyzer and kept at a distance of 5 metres from the enclosure was used to measure the strength of signal leaking out of the enclosure. Strengths of the tone were recorded for the two cases in which the door of the enclosure was kept open and subsequently closed. This differential measurement provided a measure of the isolation achievable from the enclosure. Additionally, measurements of strength of the radiated tone at increasing distances away from the enclosure showed that attenuation would increase by nearly 6 dB (Trainotti, 1990) due to propagation loss each time the distance of the receiving antenna from the enclosure was doubled. As the SARAS antenna is kept at least 100 m away from the enclosed digital receiver, the strength of any self-generated RFI from the digital receiver is suppressed by about 100 dB. The strength of emissions from the digital receiver were also measured without the shielding, and it was determined that the attenuation of the enclosure plus space loss was adequate to suppress the RFI received by the antenna to below 1 mK.

The shielded enclosure has two exhaust fans for forced cooling of the entire SARAS digital receiver system, which dissipates about 150 watts. To test the cooling efficiency of the enclosure, the digital receiver was operated in its observing configuration while two temperature probes connected to a logger recorded the ambient temperature outside the enclosure (maintained constant) and the temperature inside the enclosure. The temperature recorded by the probe inside the enclosure gradually increased from the ambient value and in about 2 hr it saturated at a value about 5 degrees C higher than the ambient value.

The pSPEC FPGA for SARAS application dissipates 12 watts. The sum of the thermal resistance of the LX240T FPGA (0.1-degree C/W) and that of the fan sink (1.7-degree C/W) mounted on the device is 1.8-degree C/W. In order to maintain the FPGA junction temperature below about 75-degree C, 10 degrees below the maximum allowed value of 85-degree C, the maximum temperature inside the RF-shielded enclosure should not exceed 41-degree C. CD/EoR observations using the SARAS radiometer is carried out only at night to avoid solar emissions and also increased terrestrial RFI in the daytime; and the ambient temperature at the observing site at night is then required to be less than 36-degree C.

## 2.5. Power Supply

As stated earlier, the precision SARAS spectral radiometer operates at long wavelengths in bands covering FM and TV channels, and is susceptible to deleterious effects of strong RFI signals from terrestrial FM, TV stations. Therefore, all sub-systems of SARAS radiometers are designed to be compact, portable and easily deployable in remote areas where the strength of terrestrial RFI is expected to be relatively small. In remote areas with no access to mains power supply, the entire radiometer depends on battery power for its operation. The SARAS digital receiver system operates off a 24 V power supply. A set of four, 12 V 100 A-hr batteries, arranged in a series-parallel combination, provides 24 V supply with adequate current capacity. The battery pack has the necessary ampere-hour capacity for at least 8 hours of continuous operation without exceeding 50% in the depth of discharge. The batteries are housed in a shielded enclosure with provision to recharge the batteries in situ, after each observing session. Two RG-400 coaxial cables are used to connect the power from the shielded enclosure containing the battery bank to the shielded enclosure housing the digital electronics: the centre conductor of one coaxial cable carries the positive 24 V supply and that of the second coaxial cable connects the negative terminal of the battery pack. Type-N connectors are used at panels and the outer conductors of both the coaxial cables are connected to the battery enclosure chassis to shield the cores carrying the positive and negative lines of the DC power supply and prevent digital noise from being radiated via the power supply cables. The power supply lines are also filtered to block RF noise.

Charging of the battery pack is done using a petrol generator or, alternately, using a set of six solar panels and a Renogy Rover MPPT Solar Charge Controller. Charging is disabled and disconnected during observing.

Using a set of DC-DC converters from Vicor Corporation, power to various sub-sections of the spectrometer are derived from the 24 V supply. A panel containing four converter modules are used to derive 19.5 V for the laptop docking station that includes a port replicator and also supplies the laptop power, a 12 V for the pSPEC card, a 5 V also for the pSPEC card and a separate 5 V for the Valon synthesizer and a Lantronix+USB based XPort card. The rubidium-disciplined crystal oscillator, PRS-10, is powered directly from 24 V.

A picture of RF-shielded enclosure containing the SARAS digital receiver is in Fig. 6.

## 3. Performance of the Spectrometer

The first SARAS radiometer deployed a fat-dipole antenna over ferrite tiles, operating usefully in the 110–175 MHz band (Patra *et al.*, 2013, 2015) and provided improved accuracy in the absolute calibration of the sky emission at 150 MHz. SARAS 2 deployed a sphere monopole antenna that operated in the 110–200 MHz band and placed the first constraints on models for CD/EoR by ruling out a subset of plausible models for thermal baryon evolution (Singh S. *et al.*, 2017, 2018b). SARAS 3, currently under development, aims at the 50–100 MHz band and targets the redshifted 21-cm absorption at Cosmic Dawn. The SARAS spectrometer is a common digital receiver for all SARAS antennas and analog receiver configurations, that may individually target up to octave bandwidth spectral segments in the 40-200 MHz band.

The signal-to-noise-and-distortion ratio (SINAD), an important dynamic performance indicator of an ADC, is defined as the ratio of the root-mean-square (RMS) signal amplitude to the mean value of the root-sum-square of noise and all other spectral components, excluding DC. In the pSPEC board, for a sampling clock of 500 MHz and for an input tone at 150 MHz with power at 1 dB below full scale ($-3$ dBm), SINAD is measured to be 48.71 dB. If we substitute this measured SINAD for SNR in Equation 1, we infer that the ENOB of the ADC in the pSPEC board is 7.8 bits, which is somewhat lower that the device datasheet specification of 8.3 bits.

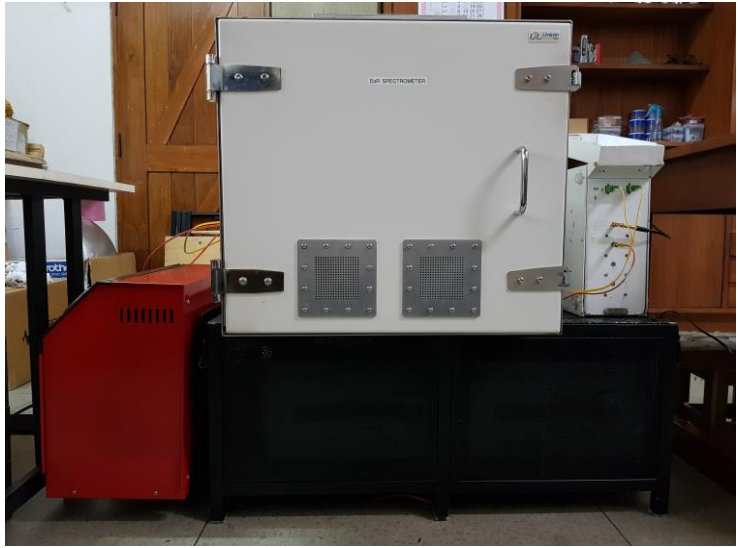

Fig. 6. The Digital Receiver is in the white colored shielded enclosure and the black painted shielded enclosure beneath the digital receiver houses the batteries. The analog signal conditioner is the unit to the right and the red unit to the left is the battery charger.

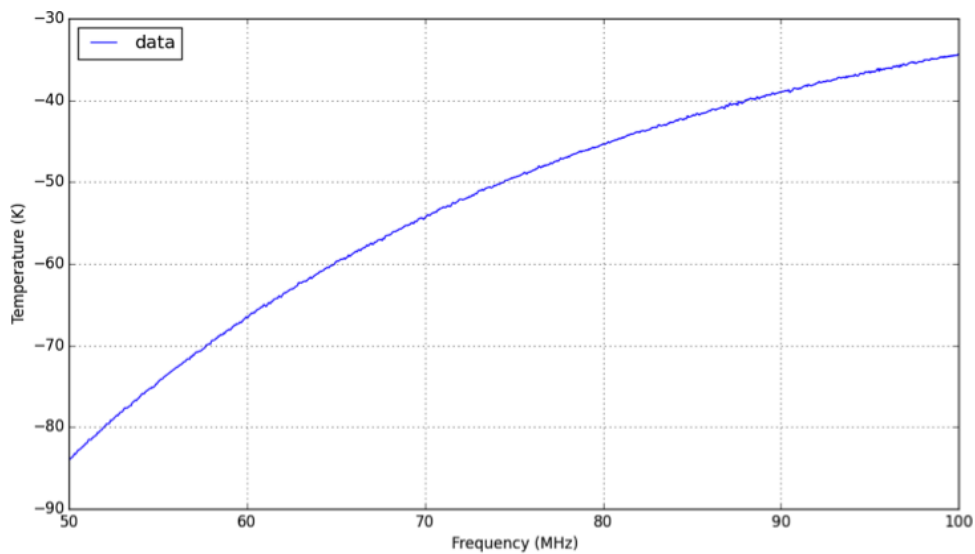

Fig. 7. Calibrated spectrum of a resistance-inductance-capacitance network as measured with the SARAS spectrometer over an octave band 50-100 MHz.

The tolerance on ADC nonlinearity is demanding because of the high dynamic range required of the spectrometer, particularly when operated in sites where significant RFI is present. A measure of ADC nonlinearity is two-tone, third-order inter-modulation distortion (IMD3). In the SARAS spectrometer, the two ADCs operate in independent channel mode without any time interleaving. For an input consisting of two closely spaced tones around 200 MHz (close to the upper edge of our band of interest), worst case IMD3 performance of about 53 dBc was measured for tone powers in the range $-16$ to $-28$ dBm.

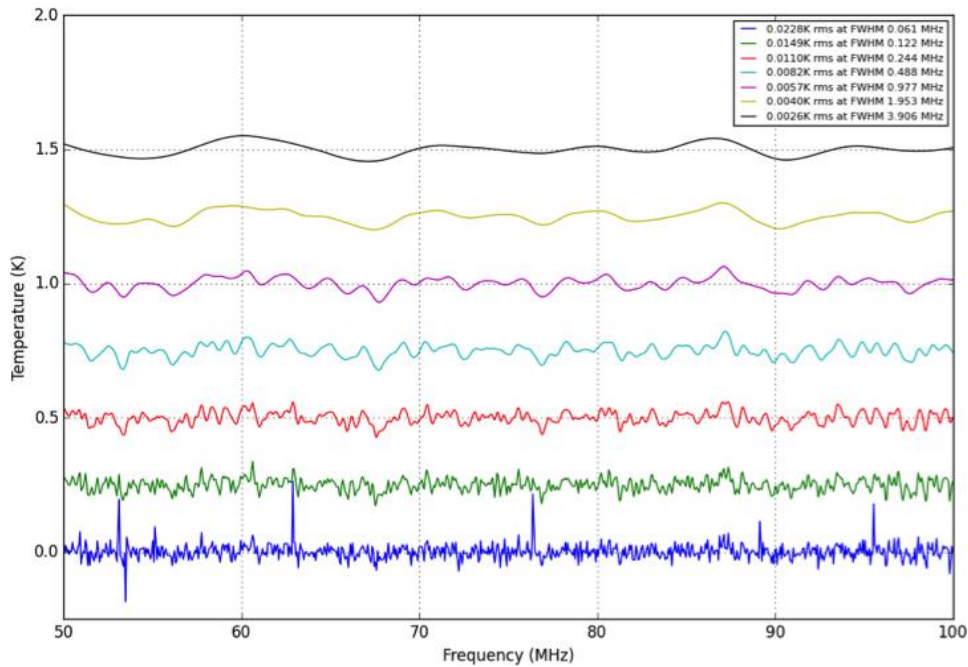

Fig. 8. Residuals on fitting out a maximally smooth polynomial from the measured spectrum. The maximum-resolution residual spectrum is shown as the lowest trace, with traces above that obtained by smoothing to progressively lower spectral resolutions. The amplitudes of the residuals have been scaled up on smoothing so that any spectral structure is apparent. The full-width at half maximum (FWHM) of the smoothing kernel and the RMS of the amplitudes of the smoothed spectra are given in the legend.

These performance measurements suggest that the SARAS spectrometer may be expected to have spurious responses at about 50 dB below the RFI. Therefore, if spurious spectral structure is to be significantly less than the expected CD/EoR signals, which is 20–200 mK, then SARAS radiometer may only be deployed at sites where RFI is significantly below 2000 K.

The performance of the spectrometer has been demonstrated in the laboratory with long duration integrations replacing the antenna with networks that provide a mismatch similar to that of a short monopole. In other words, the antenna was replaced with a resistance-inductance-capacitance network that has a frequency-dependent power reflection coefficient similar to that of the antenna. The system was operated for 17 hr, the data calibrated with switched noise injection, and the spectrum averaged. The data was calibrated for the band 50–100 MHz since the network represented a monopole that was short in this band; the measured spectrum over this octave band is shown in Fig. 7.

A maximally smooth polynomial (Sathyanarayana Rao *et al.*, 2017) was fit to the data to examine for spurious spectral structure that might confuse any detection of CD/EoR 21-cm signals. The residual to a maximally smooth fit is shown in Fig. 8. As seen from the figure, there are no spectral structures in the smoothed spectra and the RMS noise goes down as expected from a value of 22.8 mK at 61 kHz resolution to 2.6 mK at 3.9 MHz resolution. This laboratory test of the SARAS correlator performance validates its capability for detecting CD/EoR spectral features that may have amplitudes 20–200 mK.

## 4. Summary and future work

As part of the SARAS experiment that aims to detect the weak 21-cm global signal from CD/EoR, we have developed a digital correlation spectrometer system capable of processing a baseband analog signal pair with up to 250 MHz bandwidth. The RF signals from SARAS antennas are processed in analog signal chains and provided to the digital receiver as a pair of baseband analog signals; in the digital receiver the signals are digitized, then channelized using windowed FFTs to yield spectra with resolution of 61 kHz, thus producing self- and cross-power spectra of the two analog inputs. A high sidelobe suppression of 80 dB is provided by the window function; thus even if RFI as strong as $10^6$ K is present the spillover across the band into other channels is limited to be below 10 mK. The 10-bit ADC has a two-tone, third-order intermodulation performance of 53 dBc that suggests that, if unwanted spectral structure due to intermodulation distortion of closely spaced multiple RFI signals is to be less than the CD/EoR signal strength of 20–200 mK, strength of multiple RFI signals—if present—needs to be well below 2000 K. In the operating band, the RF-shielded enclosure that houses the integrated spectrometer, along with the external battery units and interconnections, provides about 75 dB attenuation of electromagnetic signals generated inside the enclosure. Since the SARAS antenna is deployed at least 100 m away from the enclosure, the combination of RF isolation provided by the enclosure and space loss ensures that at the antenna terminals RFI from the digital receiver has strength less than 1 mK. We have evaluated the system performance with long duration acquisition of data with the antenna replaced by an R-L-C network that mimics the reflection coefficient amplitude of the SARAS antenna; analysis of the data demonstrates that the digital correlation spectrometer is capable of providing spectra with spurious spectral structure well below the expected amplitudes of CD/EoR 21-cm signals.

The SARAS digital correlation spectrometer performance is currently limited primarily by the analog- to-digital converter-the sampler-owing to the limited number of quantization levels; therefore, future up- grades are to change the analog-to-digital converter to an improved device now available, with 14 bits and hence improved dynamic range and linearity performance.

**Acknowledgments**

We would like to thank Madhavi for her contributions at various stage of development of pSPEC platform like: verification of schematic, component placement, high-speed signal routing, initial power-on tests; development of filters for power supply; wiring inside the RF-shielded enclosure and incorporating modifications in the XPort card (along with Kasturi). Authors would like to thank Kamini for her contribution towards development and testing of all the DC-DC converter modules, accurate cabling and routing inside the shielded enclosure. We would like to thank Kasturi for redesigning XPort card for SARAS applications. We thank Prabu for sparing an XPort card that enabled us to get started with control & monitor inter- face to pSPEC. Santosh Harish took an active role in developing real-time data acquisition software. We also thank the Mechanical Engineering Group at RRI, led by Mohamed Ibrahim, for fabrication of trolley that houses electronic subsystems of correlation spectrometer, and a number of chassis for digital receiver. Nagaraj H. N. for quick fabrication of power supply filter cards. Shruti was extremely helpful during the measurement of RF isolation provided by the commercial RF-shielded enclosure.